\documentclass{article}
\usepackage[dvipdf]{epsfig}
\usepackage{color}
\usepackage{graphicx}

\newcommand{\ben}{\begin{eqnarray}}
 \newcommand{\een}{\end{eqnarray}}
  \def\no{\nonumber}
\def\lb{\label}

\begin{document}
\title{Fermion field as inflaton, dark energy and dark matter}

\author{Guilherme Grams\footnote{grams.guilherme@gmail.com}, {Rudinei C. de Souza}\footnote{rudijantsch@gmail.com}, {Gilberto M. Kremer}\footnote{kremer@fisica.ufpr.br}
\\ Departamento de F\'\i sica, Universidade Federal do Paran\'a,
  Curitiba, Brazil}

\date{}
\maketitle

\begin{abstract}
The search for constituents that can explain the periods of accelerating expansion of the Universe is a fundamental topic in cosmology. In this context, we investigate how fermionic fields minimally and non-minimally coupled with the gravitational field may be responsible for accelerated regimes during the evolution of the Universe. The forms of the potential and coupling of the model are determined through the technique of the Noether symmetry for two cases. The first case comprises a Universe filled only with the fermion field. Cosmological solutions are straightforwardly obtained for this case and an exponential inflation mediated by the fermion field is possible with a non-minimal coupling. The second case takes account of the contributions of radiation and baryonic matter in the presence of the fermion field. In this case the fermion field plays the role of dark energy and dark matter, and when a non-minimal coupling is allowed, it mediates a power-law inflation.
\end{abstract}


\section{Introduction}
\label{intro}
The identification of the constituents that can promote the inflationary period and the present accelerated era of the Universe is still an object of intense investigation. Generally one admits the existence of an exotic component in the Universe, with negative pressure, which is responsible for the current cosmic acceleration \cite{Zlatev, RevModPhys.75.559, Binder, RudiRaila}. The first candidate to represent this \textit{dark energy} was the cosmological constant, but other models for dark energy were also proposed. For the inflationary period, most of the models suppose a scalar field coupled with the gravitational field \cite{RevModPhys.75.559, PhysRevLett.80.1582, PhysRevD.62.063508}. Fermionic fields were also employed as candidates for the \textit{inflaton} or dark energy \cite{Sahagengrav, PhysRevD.69.124010, PhysRevD.74.124030, Marlosprd, Marlosepl, Rudifermions, fermioniccosmologies}. Important extending works concerning the fermionic fields in cosmology can be found in the references \cite{Piccongreene, Phys.Rev.D77:123535, Class.Quant.Grav.25:165014, Phys.Rev.D80:023501, Physics Letters B 718, gr-qc 1402.5880}.

The fact that common matter cannot account for the total matter of the Universe is another old problem which is still without a solution. Several candidates for \textit{dark matter} were proposed to explain this problem. In face with the problems of the dark matter and dark energy - the so-called dark sector - works have been made with two different fields to represent the dark matter and dark energy. Some of these works also admit an interaction between these fields \cite{macorra, rudicanonical, PhysRevD.79.063518}.

The aim of the present work is to describe a spatially flat homogeneous and isotropic Universe whose constituents are a fermionic field, a baryonic matter field (dust) and a radiation field (radiation and non-relativistic neutrinos). We investigate two models: i) a fermion field non-minimally coupled with the gravitational field and ii) radiation, matter (baryonic), dark matter and dark energy fields, where the dark sector and inflaton are described by the fermionic field. The first one is interesting for the study concerning the possibility that the fermion field plays the role of the inflaton. In the second one we analyse the whole evolution of the Universe, comprising its decelerated and accelerated eras (from inflation through radiation and matter eras up to the present acceleration).

The forms of the gravitational coupling and the self-interaction potential are obtained via Noether symmetry for the generic point-like Lagrangians of the two cases. Several works have already used the Noether symmetry approach to search for the forms of the coupling and potential of quintessence models \cite{PhysRevD.42.1091, PhysRevD.44.3136, Capozzielonoether, PhysRevD.80.104030, Demianskiastro, Rudifermions, Capozzielofder, Ruditaquion, Zhangyun2010, Sharif2013}. The Noether symmetry can be seen as a first principle for choosing the unknown functions of a given Lagrangian instead of an \textit{ad hoc} procedure. Our basic field equations in each case follow from the generic point-like Lagrangian corresponding to the action of each model. The resulting Dirac equations coupled with gravity and modified Friedmann equations are then solved for the couplings and potentials found through the Noether symmetry approach.

This paper is structured as follows: in the second section we present the general action from which the field equations of the two cases will be derived. In the third section we derive the Einstein and Dirac equations from the point-like Lagrangian for the first case. The Noether symmetry imposition for this case is done in subsection {3.1} and in subsection {3.2} we obtain the respective cosmological solutions. The point-like Lagrangian for the more general case and its respective field equations are presented in section {4}. The Noether symmetry analysis and the cosmological solutions for this case are done in subsection {4.1} and {4.2}, respectively. The last section is reserved for final remarks and conclusions. In this work we will adopt the metric signature $ (+,-,-,-) $ and the natural units $ 8\pi G=c=\hbar=1 $.

\section{Action}
\label{field equations}
We are interested in investigating a Universe modeled by a mixture whose constituents are the fermion field, matter and radiation. The action for a fermion field non-minimally coupled with the gravitational field reads
\ben\no
S&=&\int d^{4}x\sqrt{-g}\left\{{F(\Psi)R+\frac{i}{2}\left[\overline{\psi}\Gamma^{\mu}D_{\mu}\psi-({\overline{D}_{\mu}\overline{\psi}})\Gamma^{\mu}\psi\right]-V(\Psi)}\right\}\\\label{1}
 &+&S_{m}+S_{r},
\een
where $ R$ is the Ricci scalar, $\psi$ and $\overline \psi=\psi^\dag \gamma^0$ are the spinor field and its adjoint, respectively. In agreement to the general covariant principle  the Pauli matrices are replaced by  $ \Gamma^{\mu}=e^{\mu}_{\nu}\gamma^{\nu} $ where $e^{\mu}_{\nu}$ are tetrad fields. The generalized Dirac matrices obey the Clifford algebra $ \lbrace\Gamma^{\mu},\Gamma^{\nu}\rbrace=2g^{\mu\nu} $. $ F(\Psi) $ is the (generic) function that describes the coupling and $ V(\Psi) $ the self-interaction potential density of the fermionic field. Papers that consider $ F $ and $ V $ as functions of the bilinear $ \Psi=\overline{\psi}\psi $ are commonly found in the literature \cite{Marlosprd, Rudifermions} and the few ones that consider $ F $ and $ V $ as functions of the pseudo-scalar $ \Psi=\overline{\psi}\gamma^{5}\psi $ do it in an \textit{ad hoc} way \cite{Sahagengrav, PhysRevD.69.124010, Marlosprd}. Here we shall consider that $ F $ and $ V $ are only functions of the pseudo-scalar $ \Psi=\overline{\psi}\gamma^{5}\psi $ which will be found by the Noether symmetry criterion. Furthermore, $ S_m $ is the action of the matter field and $ S_r $ the action of the radiation field. The covariant derivatives in (\ref{1}) read
\ben\label{1a}
D_\mu\psi= \partial_\mu\psi-\Omega_\mu\psi,\qquad
D_\mu\overline\psi=\partial_\mu\overline\psi+\overline\psi\Omega_\mu,\\\Omega_\mu=-\frac{1}{4}g_{\rho\sigma}\left[\Gamma^\rho_{\mu\delta}
-e_b^\rho\left(\partial_\mu e_\delta^b\right)\right]\Gamma^\delta\Gamma^\sigma.
\een
Here $\Omega_\mu$ denotes  the spin connection
and $\Gamma^\nu_{\sigma\lambda}$  the Christoffel symbols.

We will firstly analyse the  case where the fermion field non-minimally couples with the gravitational field in the absence of matter and radiation, i.e., $ S_m =0 $ and $ S_r = 0$.

\section{Field equations for the fermion field}

For a spatially flat Friedmann-Robertson-Walker metric $ds^2=dt^2-a(t)^2(dx^2+dy^2+dz^2)$ --
where $a(t)$ denotes the cosmological scale factor -- the  Dirac-Pauli matrices and spin connection become
\begin{equation}\label{1c}
\Gamma^0=\gamma^0,\qquad
\Gamma^i=\frac{1}{a(t)}\gamma^i, \qquad \Omega_0=0, \qquad\Omega_i=\frac{1}{2}\dot a(t)\gamma^i\gamma^0,
\end{equation}
with the dot denoting time derivative.

In this case we can obtain -- after a partial integration of the action (\ref{1}) (with $ S_m =0 $ and $ S_r = 0$) -- the point-like Lagrangian
\begin{equation}\label{2}
\mathcal{L} =6a\dot{a}^{2}F+6a^{2}\dot{a}\dot{\Psi}F^{\prime}+a^{3}\frac{i}{2}\left(\dot{\overline{\psi}}\gamma^{0}\psi-\overline{\psi}\gamma^{0}\dot{\psi}\right)+a^{3}V.
\end{equation}
Here the derivative with respect to the pseudo-scalar $ \Psi $ is represented by a prime.

The Dirac equations for the spinor field and its adjoint follow from the Euler-Lagrange equations for $ \psi $ and $ \overline{\psi} $ applied to the Lagrangian (\ref{2}), namely,
\ben\label{3}
\dot{\overline{\psi}}+\frac{3}{2}H\overline{\psi}-iV^{\prime}\overline{\psi}\gamma^{5}\gamma^{0}+6iF^{\prime}\overline{\psi}\gamma^{5}\gamma^{0}\left(\dot{H}+2H^{2}\right)=0,
\\\label{4}
\dot{\psi}+\frac{3}{2}H{\psi}+iV^{\prime}\gamma^{0}\gamma^{5}{\psi}-6iF^{\prime}\gamma^{0}\gamma^{5}{\psi}\left(\dot{H}+2H^{2}\right)=0,
\een
where $ H=\dot{a}/a $ denotes de Hubble parameter.

From the Euler-Lagrange equations for $ a $ applied to the Lagrangian (\ref{2}) we obtain the acceleration equation
\begin{equation}\label{5}
\frac{\ddot{a}}{a}=-\frac{\rho_{f}+3p_{f}}{12F},
\end{equation}
where the energy density and pressure of the fermion field, read
\ben\label{6a}
\rho_{f}=V-6HF^{\prime}\dot{\Psi},\\\lb{6b}
p_{f}=\left[V^{\prime}-6F^{\prime}(\dot{H}+2H^{2})\right]\Psi -V+2\left(\ddot{\Psi}F^{\prime}+\dot{\Psi}^{2}F^{\prime\prime}+2\dot{\Psi}F^{\prime}H\right).
\een

Finally by imposing that the energy function associated with the Lagrangian (\ref{2}) vanishes:
\ben\label{8}
E_{\mathcal{L}}=\frac{\partial \mathcal{L}}{\partial \dot{a}}\dot{a}+\dot{\overline{\psi}}\frac{\partial \mathcal{L}}{\partial \dot{\overline{\psi}}}+\frac{\partial \mathcal{L}}{\partial \dot{\psi}}\dot{\psi}-\mathcal{L}=0,
\een
we get the Friedmann equation
\begin{equation}\label{9}
H^{2}=\frac{V-6HF^{\prime}\dot{\Psi}}{6F}=\frac{\rho_{f}}{6F}.
\end{equation}

Note that we supposed that the spinors which are only functions of time are compatible with the homogeneity and isotropy of a flat Friedmann-Robertson-Walker metric. In principle, this can be achieved for the classical spinors we are considering once the corresponding energy-momentum tensor is not anisotropic, so that the description of the energy density and pressure of the fluid is consistent with this metric. However, even before a semiclassical quantization of the spinor field in the presence of gravity, the ansatz for fermions (classical) compatible with the Friedmann-Robertson-Walker background is not a trivial subject. The marriage of the fermions and homogeneous and isotropic geometry of general relativity is discussed by references \cite{Christodoulakis, Christodoulakis1} in the context of Dirac fields concerning the classical field equations and Dirac's formalism for constrained Hamiltonian systems. The corresponding results appear to be in favor of the compatibility of the ansatz $\psi(t)$ with a homogeneous and isotropic Universe. The mentioned references are on the long road of investigation on the issue of quantum fields living in the space-time of general relativity, focused on a semiclassical approach in the cosmological context. Former papers which attacked this problem can be found in references \cite{Nelson, Nelson1}, as well as an extension of these approaches can be seen in \cite{D'Eath}. It is important to point out that the effective behavior of the cosmological fluid is mathematically described by a classical spinor field, which is not necessarily saying that the fluid is fundamentally composed by fermionic particles which are likely to be detected. In principle, classical spinor fields do not have anything to do with real fermionic particles. In fact, physically speaking, fermionic particles only exist in the quantum and relativistic level, whose marriage produces the known Dirac equation. The same can be said when describing inflation with a classical scalar field. Such a field models the effective behavior of an exotic fluid, but this not necessarily mean that the fluid is composed by unknown (or known) bosonic particles. A clear discussion on cosmological applications of classical fermion fields can be also found in the work \cite{Magueijo}.

\subsection{Noether symmetry}

In terms of the components of the spinor field, $ \psi=(\psi_{1},\psi_{2},\psi_{3},\psi_{4})^{T} $ and its adjoint $ \overline{\psi}=(\psi^{\dagger}_{1},\psi^{\dagger}_{2},-\psi^{\dagger}_{3},-\psi^{\dagger}_{4}) $, the Lagrangian (\ref{2}) can be written as
\ben\no
 \mathcal{L}=6a\dot{a}^{2}F+6a^{2}\dot{a}F^{\prime}\sum_{i,j=1}^4 \left(\dot{\psi}^{\dagger}_{i}\psi_{j}+\psi^{\dagger}_{i}\dot{\psi_{j}}\right)\varepsilon_{ij}
\\\label{10}
+\frac{i}{2}a^{3}\sum_{i=1}^4 \left(\dot{\psi}^{\dagger}_{i}\psi_{i}-\psi^{\dagger}_{i}\dot{\psi_{i}}\right)+a^{3}V,
\een
which is only function of ($ a $, $ \dot{a}$, $\psi^{\dagger}_{l} $, $ \dot{\psi}^{\dagger}_{l} $, $\psi_{l} $, $\dot{\psi}$). In the Lagrangian (\ref{10}) it was introduced the symbol $ \varepsilon_{ij} $, which assumes the values
 \begin{equation}\label{11}
\left\{
  \begin{array}{lllll}
    \varepsilon_{ij}=-1  & \hbox{for} &\varepsilon_{13} &\hbox{or}  &\varepsilon_{24}, \\
    \varepsilon_{ij}=+1 & \textrm{for} & \varepsilon_{31} & \textrm{or} &\varepsilon_{42}, \\
   \varepsilon_{ij}=0 &  \textrm{otherwise.} & & &
  \end{array}
\right.
  \end{equation}

The Noether symmetry is satisfied by the condition (see Appendix)
\begin{equation}\label{12}
L_{\textbf{X}}\mathcal{L}=\textbf{X}\mathcal{L}=0,
\end{equation}
where \textbf{X} is the infinitesimal generator of symmetry defined by
\begin{equation}\label{13}
\textbf{X}=C_0\frac{\partial}{\partial a}+\dot{C_0}\frac{\partial}{\partial \dot{a}}+\sum_{l=1}^{4}\left(C_{l}\frac{\partial}{\partial \psi_{l}^{\dagger}}+D_l\frac{\partial}{\partial \psi_l}+\dot{C_l}\frac{\partial}{\partial \dot{\psi_{l}}^{\dagger}}+\dot{D_l}\frac{\partial}{\partial \dot{\psi_l}}\right),
\end{equation}
and $ L_{\textbf{X}} $ is the Lie derivative of $ \mathcal{L}$ with respect to the vector \textbf{X} which is defined in the tangent space. Furthermore, the parameters $ C_0 $, $ C_l $ and $ D_l $ are arbitrary functions of ($ a $, $ \psi_{l} $, $ \psi^{\dagger}_{l} $).

Applying the condition (\ref{12}) to the Lagrangian (\ref{10}), one has an equation whose terms depend explicitly on $ \dot{a}\dot{\psi}_l^{\dagger} $, $ \dot{a}\dot{\psi}_l $, $ \dot{a}^2 $, $ \dot{\psi}_l^{\dagger} $, $ \dot{\psi}_l $, $ \dot{\psi}_l^{\dagger}\dot{\psi}_m $, $ \dot{\psi}^{\dagger}_l\dot{\psi}^{\dagger}_m $, $ \dot{\psi}_l\dot{\psi}_m $ and $ \dot{a} $. By equating the coefficients of the referred terms to zero, we obtain the following system of coupled differential equations
\ben\no
 F^{\prime}\psi_j\varepsilon_{lj}\left(2C_0+a\frac{\partial C_{0}}{\partial a}\right)+2F\frac{\partial C_0}{\partial \psi^{\dagger}_l}+aF^{\prime\prime}\psi_j\varepsilon_{lj}\sum_{i,k=1}^{4}\left(C_i\psi_k+\psi^{\dagger}_{i}D_k\right)\varepsilon_{ik}
\\\label{14}+aF^{\prime}\left[D_{j}\varepsilon_{lj}+\sum_{i,k=1}^{4}\left(\frac{\partial C_i}{\partial \psi_{l}^{\dagger}}\psi_k+\psi^{\dagger}_{i}\frac{\partial D_k}{\partial \psi_{l}^{\dagger}}\right)\varepsilon_{ik}\right]=0,
\\\no
 F^{\prime}\psi^{\dagger}_{j}\varepsilon_{jl}\left(2C_{0}+a\frac{\partial C_{0}}{\partial a}\right)+2F\frac{\partial C_{0}}{\partial \psi_{l}}+aF^{\prime\prime}\psi^{\dagger}_{j}\varepsilon_{jl}\sum_{i,k=1}^{4}\left(C_{i}\psi_{k}+\psi^{\dagger}_{i}D_{k}\right)\varepsilon_{ik}
\\\label{15}
+aF^{\prime}\left[C_{j}\varepsilon_{jl}+\sum_{i,k=1}^{4}\left(\frac{\partial C_{i}}{\partial \psi_{l}}\psi_{k}+\psi^{\dagger}_{i}\frac{\partial D_k}{\partial \psi_l}\right)\varepsilon_{ik}\right]=0,
\\\no
C_{0}F+2aF\frac{\partial C_0}{\partial a}+aF^{\prime}\sum_{i,j=1}^{4}\left(C_{i}\psi_{j}+\psi_{i}^{\dagger}D_{j}\right)\varepsilon_{ij}
\\\label{16}
+a^{2}F^{\prime}\sum_{i,j=1}^{4}\left(\frac{\partial C_i}{\partial a}\psi_j+\psi^{\dagger}_{i}\frac{\partial D_j}{\partial a}\right)\varepsilon_{ij}=0,
\\\label{17}
3C_{0}\psi_l+aD_l+a\sum_{i=1}^{4}\left(\psi_{i}\frac{\partial C_i}{\partial \psi^{\dagger}_{l}}-\psi^{\dagger}_{i}\frac{\partial D_i}{\partial \psi^{\dagger}_{l}}\right)=0,
\\\label{18}
3C_{0}\psi^{\dagger}_{l}+aC_l-a\sum_{i=1}^{4}\left(\psi_{i}\frac{\partial C_i}{\partial \psi_{l}}-\psi^{\dagger}_{i}\frac{\partial D_i}{\partial \psi_{l}}\right)=0,
\\\label{19}
F^{\prime}\left(\psi^{\dagger}_{i}\frac{\partial C_0}{\partial \psi^{\dagger}_{j}}+\psi_{j}\frac{\partial C_0}{\partial \psi_i}\right)\varepsilon_{ij}=0,\qquad F^{\prime}\left(\psi_{j}\frac{\partial C_0}{\partial \psi^{\dagger}_{m}}+\psi_{m}\frac{\partial C_0}{\partial \psi^{\dagger}_{j}}\right)\varepsilon_{ij}=0,
\\\label{20}
F^{\prime}\left(\psi^{\dagger}_{i}\frac{\partial C_0}{\partial \psi_{l}}+\psi^{\dagger}_{l}\frac{\partial C_0}{\partial \psi_{i}}\right)\varepsilon_{ij}=0,\qquad \sum_{i=1}^{4}\left(\psi_{i}\frac{\partial C_i}{\partial a}-\psi^{\dagger}_{i}\frac{\partial D_i}{\partial a}\right)=0.
\een

There remains an equality that involves the potential density, namely,
\begin{equation}\label{21}
3C_{0}V+aV^{\prime}\sum_{i,j=1}^{4}(C_{i}\psi_{j}+\psi^{\dagger}_{i}D_j)\varepsilon_{ij}=0.
\end{equation}

The above system with 55 differential equations, (\ref{14}) through (\ref{21}), will be examined in the following. From (\ref{19}) and $ (\ref{20})_{1} $ we infer that there are two possibilities for the coupling, $ F^{\prime}=0 $ and $ F^{\prime}\neq0 $. These two possibilities will be analyzed separately.
	
\textbf{(1) Case} $ F'=0 $:

If $ F'=0 $ it follows that $ F= $ constant and equations (\ref{19}) and $ (\ref{20})_{1} $ are automatically satisfied. Furthermore, from equations (\ref{14}) and (\ref{15}) we have that $ C_0 = C_0 (a)$. Hence, equation (\ref{16}) determines the form of $C_0$, namely,
$C_{0}=\lambda/a^{1/2}$,
where $ \lambda $ is a constant. Equations $ (\ref{20})_{2} $ and (\ref{17})  furnish an expression for $ C_j $ and $ D_j $
\begin{equation}\label{23}
C_j=-\frac{3}{2}\frac{\lambda\psi^{\dagger}_{j}}{a^{3/2}}+\beta\psi^{\dagger}_{i}\varepsilon_{ij},\qquad D_j=-\frac{3}{2}\frac{\lambda\psi_{j}}{a^{3/2}}+\beta\psi_{i}\varepsilon_{ij},
\end{equation}	
with $ \beta $ being a constant.

Finally, from (\ref{21}) we determine the potential
\begin{equation}\label{24}
V=\alpha\Psi,
\end{equation}
where $ \alpha $ is a constant.

\textbf{(2) Case} $ F'\neq 0 $:

Now we will analyse the case where $ F $ can be an arbitrary function of the pseudo-scalar $ \Psi $. For this end, we write equation (\ref{21}) as follows
\begin{equation}\label{25}
\sum_{i,j=1}^{4}(C_{i}\psi_{j}+\psi^{\dagger}_{i}D_j)\varepsilon_{ij}=-3\frac{C_0}{a}\frac{V}{V^{\prime}}.
\end{equation}
By the differentiation of the above equation with respect to $ a $, one has
\begin{equation}\label{26}
\sum_{i,j=1}^{4}\left(\frac{\partial C_{i}}{\partial a}\psi_{j}+\psi^{\dagger}_{i}\frac{\partial D_j}{\partial a}\right)\varepsilon_{ij}=\frac{3}{a}\frac{V}{V^{\prime}}\left(\frac{C_0}{a}-\frac{\partial C_0}{\partial a}\right).
\end{equation}

	Now we insert equations (\ref{25}) and (\ref{26}) into (\ref{16}) and, recalling that $ F $ and $ V $ are only functions of $ \Psi $, the corresponding result is
\begin{equation}\label{27}
\frac{a}{C_0}\frac{\partial C_0}{\partial a}=\frac{FV^{\prime}}{3F^{\prime}V-2FV^{\prime}}=k,
\end{equation}	
	where $ k $ is a constant.
	
	By analyzing equations (\ref{19}) and $ (\ref{20})_{1} $ with $ F^{\prime}\neq0 $, we infer that $ C_0 $ does not depend on $ \psi^{\dagger}_{j} $ and $ \psi_{j} $, i.e., $ C_0=C_{0}(a) $. Then, from equation (\ref{27}) we obtain the solution $C_0=\lambda a^{k}$, where $ \lambda $ is a constant.

	By combining equations $ (\ref{20})_{2} $ with (\ref{17})  we find the expressions for $ C_j $ and $ D_j $, respectively,
\begin{equation}\label{29}
C_j=-\frac{3}{2}\lambda a^{k-1}\psi^{\dagger}_{j}+\beta\psi^{\dagger}_{i}\varepsilon_{ij}\qquad D_j=-\frac{3}{2}\lambda a^{k-1}\psi_{j}+\beta\psi_{i}\varepsilon_{ij}.
\end{equation}

	From equation (\ref{21}) it follows that the potential has the same linear form obtained in the case for $F'=0$, namely, $V=\alpha\Psi.$ From equations (\ref{24}) and (\ref{27}) we obtain a power-law function for the coupling, $ F=\omega \Psi ^{p} $, where $ \omega $ is a constant. The exponent $p$ of the power-law coupling results from equations (\ref{15}), (\ref{16}) and  (\ref{27}) and must satisfy the following relationships
 \ben\label{30}
\left\{
  \begin{array}{ll}
    3pk=1+2k, \\
   3p=2+k,
  \end{array}
\right.
\qquad\hbox{which implies}\qquad
\left\{
  \begin{array}{ll}
    (1,1), \\
    (-1,1/3),
  \end{array}
\right.
\een
from which we determined $p$ and $ k $.
	
	Hence, we conclude that for $ F'\neq 0$ there are two Noether symmetries, namely,

\textit{Symmetry 1} (for $ k=1 $ and $ p=1 $):

\ben
F=\omega\Psi,\qquad V=\alpha\Psi,\qquad C_{0}=\lambda a, \\
C_j=-\frac{3}{2}\lambda\psi^{\dagger}_{j}+\beta\psi^{\dagger}_{i}\varepsilon_{ij}, \qquad D_j=-\frac{3}{2}\lambda\psi_{j}+\beta\psi_{i}\varepsilon_{ij}.
\een

\textit{Symmetry 2} (for $ k=-1 $ and $ p=1/3 $):

\ben
 F=\omega\Psi^{1/3},\qquad V=\alpha\Psi,\qquad C_{0}=\frac{\lambda}{a},\\ C_j=-\frac{3}{2}\frac{\lambda\psi^{\dagger}_{j}}{a^{2}}+\beta\psi^{\dagger}_{i}\varepsilon_{ij}, \qquad D_j=-\frac{3}{2}\frac{\lambda\psi_{j}}{a^{2}}+\beta\psi_{i}\varepsilon_{ij}.
\een

\subsection{Cosmological solutions}

 From the Dirac equations of the spinor field and its adjoint coupled with the gravitational field, (\ref{3}) and (\ref{4}), we obtain a differential equation for $\Psi $,
 \begin{equation}\label{31}
\frac{d\Psi}{dt}+3\frac{\Psi}{a}\frac{da}{dt}=0,\;\;\;\;\; \textrm{so that,}\;\;\;\;\;  \Psi=\frac{\Psi_0}{a^{3}},
\end{equation}
	where $ \Psi_0 $ is a constant.
	
	From this result, we are able to solve the Friedmann equation and determine the time evolution of the scale factor. As we did before, the two cases $ F'=0 $ and $F'\neq0 $ will be analyzed separately.

\subsubsection{The case $F'=0$:}

The choice $F=$ constant	$=1/2$ refers to a minimal coupling of the fermionic field with the gravitational field. Hence, the Friedmann equation (\ref{9}) leads to a power-law solution for the time evolution of the scale factor, i.e.,
\begin{equation}\label{32}
a(t)=[\Omega(t-t_{0})]^{2/3},\qquad \textrm{with,}\qquad \Omega=\frac{3}{2}\sqrt{\frac{\alpha\Psi_{0}}{3}}.
\end{equation}
Such a solution describes a decelerated Universe with a matter dominated behavior.

Equations (\ref{6a}) and (\ref{6b}) furnish the energy density and pressure of the fermion field, namely,
\begin{equation}\label{33}
\rho_{f}=\frac{\alpha\Psi_0}{a^{3}},\qquad p_{f}=0.
\end{equation}
From the above expressions we conclude that the $ F'=0$ case describes a pressureless matter field.\\

\subsubsection{The case $F'\neq0$:}

Solving the Friedmann equation (\ref{9}) for the \textit{Symmetry 1}, we have the following solution for the scale factor
\begin{equation}\label{34}
a(t)=e^{\Lambda(t-t_{0})},\;\;\;\; \textrm{with,}\;\;\;\; \Lambda=\sqrt{\frac{-\alpha}{12\omega}}.
\end{equation}
The above exponential solution shows that the fermion field plays the role of an inflaton.

The corresponding energy density and pressure now read
\begin{equation}\label{35}
\rho_f=-\frac{\alpha\Psi_0}{2a^{3}},\;\;\;\;\;\;\;\;\;\;p_f=-\rho_f .
\end{equation}
	
	From the weak energy condition we know that the energy density is  always a positive quantity \cite{hawking}. Hence, we infer from $ (\ref{35})_1 $ that $ \alpha \Psi_{0} < 0 $. Since the coupling has to be a positive quantity, we should have $ \omega \Psi_0 > 0 $, i.e., the constants $ \alpha $ and $ \omega $ always will have opposite sings, which ensures a positive real value for $ \Lambda $. Furthermore, we can infer form equation $ (\ref{35})_2 $ that the pressure of the fermion field is always negative and proportional to the energy density.
	
 In this exponential scenario, we get from equation $ (\ref{31})_2 $ the following solution for the time evolution of $ \Psi $
\begin{equation}\label{36}
\Psi (t)=\Psi_{0}e^{-3\Lambda(t-t_0)}.
\end{equation}

Now one can also determine the time evolution of the energy density and pressure of the fermion field, which reads
\begin{equation}\label{37}
p_f (t)=-6\omega\Lambda^{2}\Psi_{0}e^{-3\Lambda(t-t_0)}=-\rho_f (t).
\end{equation}

Although equation (\ref{34}) indicates an eternal accelerated expansion, equations (\ref{36}) and (\ref{37}) show that the source of this expansion should come to an end, since the energy and potential densities of the fermion field tend to zero. Then another regime should take the place of the exponential phase.

For the \textit{Symmetry 2} the Friedmann equation does not have a solution.

\section{Field equations for the fermion, radiation and matter fields}

Now we will look at the action with the fermion, matter and radiation fields contributions. By integrating by parts the action (\ref{1}) for a spatially flat Friedmann-Robertson-Walker metric, we have the point-like Lagrangian
\begin{equation}\label{38}
\mathcal{L} =6a\dot{a}^{2}F+6a^{2}\dot{a}\dot{\Psi}F^{\prime}+a^{3}\frac{i}{2}\left(\dot{\overline{\psi}}\gamma^{0}\psi-\overline{\psi}\gamma^{0}\dot{\psi}\right)+a^{3}V+\rho^{0}_{m}+\frac{\rho^{0}_{r}}{a},
\end{equation}
where $ \rho^{0}_{m} $ and $ \rho^{0}_{r}$ are the energy density of the baryonic matter and radiation at a initial instant $ t=t_{0} $, respectively.

Note that the new Lagrangian (\ref{38}) does not present extra terms with respect to the spinor field in comparison to the Lagrangian (\ref{2}). Hence, the Dirac equations are the same as those given by (\ref{3}) and (\ref{4}). However, by imposing that the energy function associated with the Lagrangian (\ref{38}) vanishes, we obtain a different Friedmann equation
\begin{equation}\label{39}
H^{2}=\frac{\rho_{T}}{6F},\qquad\hbox{where}\qquad \rho_{T}=V-6HF'\dot{\Psi}+\frac{\rho^{0}_{m}}{a^{3}}+\frac{\rho^{0}_{r}}{a^{4}}
\end{equation}
 stands for the total energy density, which includes the energy densities of the matter and radiation fields. In the acceleration equation (\ref{5}) one has only to include the radiation pressure $p_r=\rho_r/3$, since the matter field is supposed to be pressureless.

\subsection{Noether symmetry}

  As we did before we rewrite the Lagrangian (\ref{38}) in terms of the components of the spinors as
\ben\no
 \mathcal{L}=6a\dot{a}^{2}F+6a^{2}\dot{a}F^{\prime}\sum_{i,j=1}^4 \left(\dot{\psi}^{\dagger}_{i}\psi_{j}+\psi^{\dagger}_{i}\dot{\psi_{j}}\right)\varepsilon_{ij}
\\\label{41}
+\frac{i}{2}a^{3}\sum_{i=1}^4 \left(\dot{\psi}^{\dagger}_{i}\psi_{i}-\psi^{\dagger}_{i}\dot{\psi_{i}}\right)+a^{3}V+\rho^{0}_{m}+\frac{\rho^{0}_{r}}{a}.
\een

 By imposing the Noether symmetry to the above Lagrangian and performing the same analysis as in previous section, we get the same system of coupled differential equations (\ref{14}) -- (\ref{20}), except for the differential equation which involves  the potential. Such an equation now reads
\begin{equation}\label{42}
C_{0}\left(3V-\frac{\rho^{0}_{r}}{a^{4}}\right)+aV^{\prime}\sum_{i,j=1}^{4}(C_{i}\psi_{j}+\psi^{\dagger}_{i}D_j)\varepsilon_{ij}=0.
\end{equation}

As in the previous section, we infer from  (\ref{19}) and $ (\ref{20})_{1} $ that there are two possibilities for the coupling, $ F^{\prime}=0 $ and $ F^{\prime}\neq0 $.

If $ F^{\prime}=0 $ it follows that $ F=$ constant and equations (\ref{19}) and $ (\ref{20})_{1} $ are identically satisfied. Equation (\ref{42}) is solved for $ C_{0}=0$ and $ V=$ constant or $ C_{i}=-f(a)\psi^{\dagger}_{i} $ and $ D_{j}=f(a)\psi_{j} $, with $ f(a) $ being an arbitrary function of $ a $. With $ C_{0}=0$ and $ F^{\prime}=0 $ equations (\ref{14}) -- (\ref{16}) are also satisfied as well as (\ref{17})  with $ C_{i}=-f(a)\psi^{\dagger}_{i} $ and $ D_{j}=f(a)\psi_{j} $. From $ (\ref{20})_{2} $ it follows that $ f(a)=\gamma $, where $ \gamma $ is a constant, and
$C_{l}=- \gamma \psi^{\dagger}_{l}$, $D_{l}= \gamma \psi_{l}$.

We note from the above results that  (\ref{42}) is also satisfied by an arbitrary potential $ V=V(\Psi) $, including $V=$ constant.
Observe that the system (\ref{14}) -- (\ref{20}) and (\ref{42}) are solved with the same symmetry for $ F^{\prime}= 0 $ and $ F^{\prime}\neq 0 $ (i.e., for an arbitrary $F$, including $F=$ constant).

Then for $C_{0}=0$, $C_l=- \gamma \psi^{\dagger}_{l}$ and $D_l=\gamma \psi_{l}$ the self-interaction potential of the fermion field is an arbitrary function of the pseudo scalar, when the
 coupling function is a constant or it is an arbitrary function of the pseudo scalar, i.e.,
\begin{equation}
\left\{
  \begin{array}{ll}
    F=\hbox{constant}\Longrightarrow V=V(\Psi), \\
    F=F(\Psi)\Longrightarrow V=V(\Psi).
  \end{array}
\right.
\end{equation}

\subsection{Cosmological solutions}	
\subsubsection{The case $F'=0$:}	
	The solution for the pseudo-scalar is the same as that in the previous section, namely, $ \Psi=\Psi_{0}/a^{3} $ and with $F=$ constant $=1/2 $ the fermion field is minimally coupled with the gravitational field. For the potential density we propose a linear combination of two powers of the pseudo-scalar
\begin{equation}
V(\Psi)=V_{0}\Psi ^{n} + \Lambda \Psi ^{l},
\end{equation}
where $ V_{0} $, $ \Lambda $, $ n $ and $ l $ are free parameters. Here we are working with the case where we have the contributions of matter and radiation. For a more complete model of the present status of the Universe we need the contributions of dark matter and dark energy. In this way, we may take the adequate powers as $ n=1 $ and $ l=0 $. Thus, as the energy density of the fermionic model for a minimally coupled field   according to (\ref{6a}) is $ \rho_{f}=V $, it follows that
\begin{equation}\lb{pot}
\rho_{f}=V_{0}\Psi + \Lambda =\rho_{dm}+\rho_{de}
\end{equation}
where $ \rho_{dm}=V_{0}\Psi_0/a^3 $ and $ \rho_{de}=\Lambda $ represent the energy density of the dark matter and dark energy, respectively.

The Friedmann equation for this case does not have an analytic solution for the time evolution of the scale factor. So we perform a change of variables to analyse the cosmological behavior of the proposed model. The red-shift will be used as a variable instead of time thanks to the following relationships:	
\begin{equation}
\frac{1}{a}=z+1, \qquad \frac{d}{dt}=-H(1+z)\frac{d}{dz}.
\end{equation}

Afterwards, the total energy density and the total pressure can be expressed in terms of the red-shift, yielding,
\begin{equation}
\rho_{T} (z)= (V_{0}\Psi_0+\rho^{0}_{m})(z+1)^3 +\rho_{r}^{0}(z+1)^4 +\Lambda,
\quad
p_T (z) =\frac{1}{3}\rho^{0}_{r}(1+z)^{4}-\Lambda.\quad
\end{equation}

In order to compare our results with the observational data, we divide all the above equations by $ \rho_0 $ -- the critic density when $ z=0 $ -- from which one has the following expressions:
  \begin{equation}
\overline{\rho_{T}(z)}=\overline{\rho_{dm}}+\overline{\rho_{m}}+\overline{\rho_{r}}+\overline{\rho_{de}},
\qquad  \overline{p_{T}(z)}=\frac{1}{3}\overline{\rho^{0}_{r}}(z+1)^{4}-\overline{\rho_{de}},
  \end{equation}
where the bar indicates that the quantity was divided by $ \rho _{0} $ and
\begin{equation}
\overline{\rho_{m}}=\overline{\rho_{m}^{0}}(z+1)^3,\quad\overline{\rho_{r}} =\overline{\rho_{r}^{0}}(z+1)^4,
\quad
\overline{\rho_{dm}} =\overline{V_0\Psi_0}(z+1)^3,\quad \overline{\rho_{de}} =\overline{\Lambda}.\quad
\end{equation}

The plots of the density parameter were done with the initial conditions which match the astronomical data. At $z=0$ we introduce the quantities $ \overline{\rho_{m}}(0)=\rho^{0}_{m}/ \rho_{0}=\Omega^{0}_{m} $, $ \overline{\rho_{r}}(0)=\rho^{0}_{r}/ \rho_{0}=\Omega^{0}_{r} $, $ \overline{\rho_{dm}}(0)=\rho^{0}_{dm}/ \rho_{0}=\Omega^{0}_{dm} $ and $ \overline{\rho_{de}}(0)=\rho^{0}_{de}/ \rho_{0}=\Omega^{0}_{de} $, where $\Omega^{0}_{i}$ denotes the value of the density parameter of each component at present time. The values adopted here are:  $\Omega^{0}_{m}=0.0463$, $\Omega^{0}_{dm}=0.233$, $\Omega^{0}_{de}=0.721$ and, $\Omega^{0}_{r}=8.5\times 10^{-5}$ (see \textit{e.g.} \cite{fukugita2004, WMAP9}).
\begin{figure}
\begin{minipage}[b]{0.5\linewidth} 
\centering
\includegraphics[width=7cm]{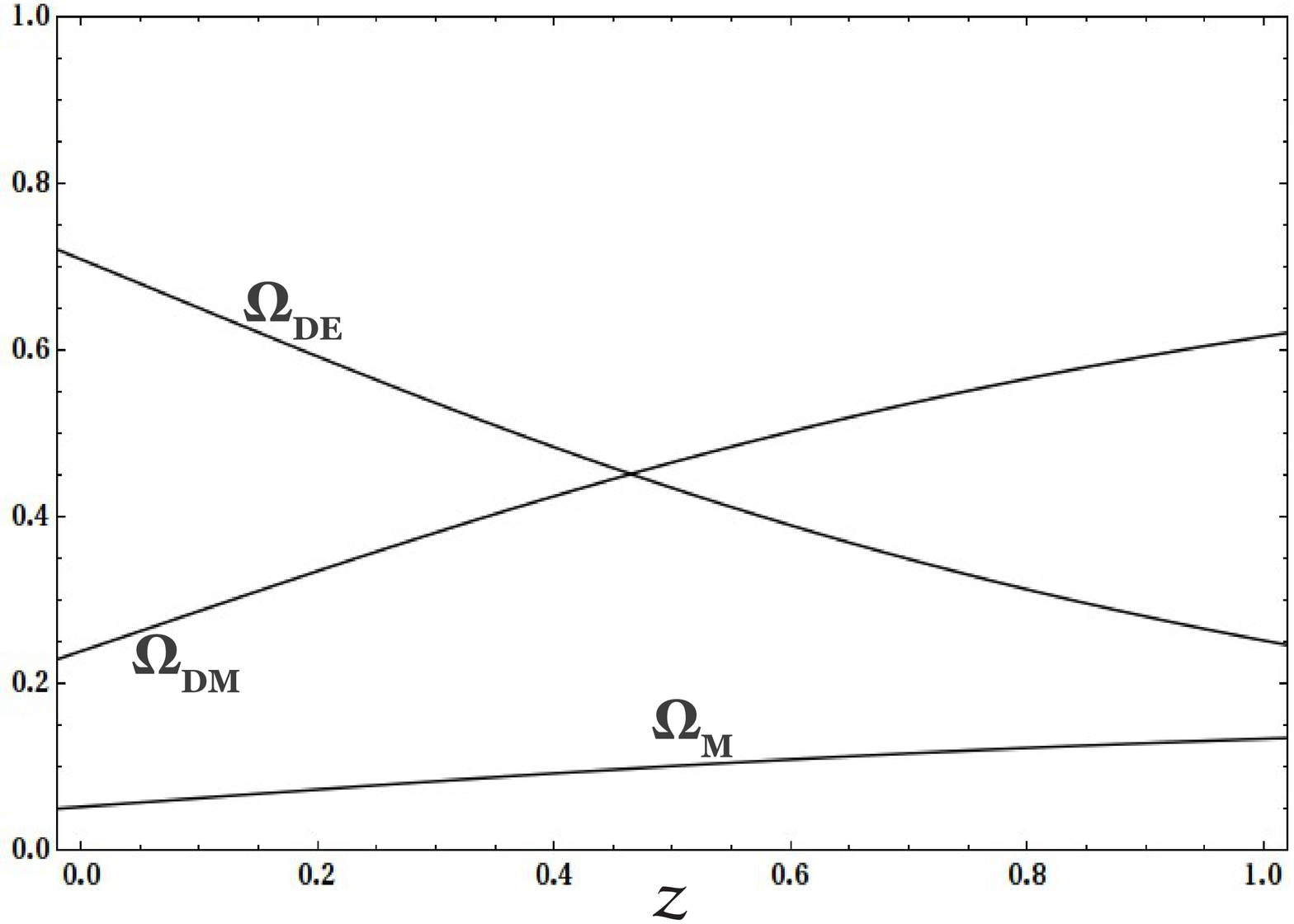}
\caption{Density parameters for \- $ 0\leq z \leq 1 $.}
\end{minipage}
\hspace{0.3cm} 
\begin{minipage}[b]{0.5\linewidth}
\centering
\includegraphics[width=7.05cm]{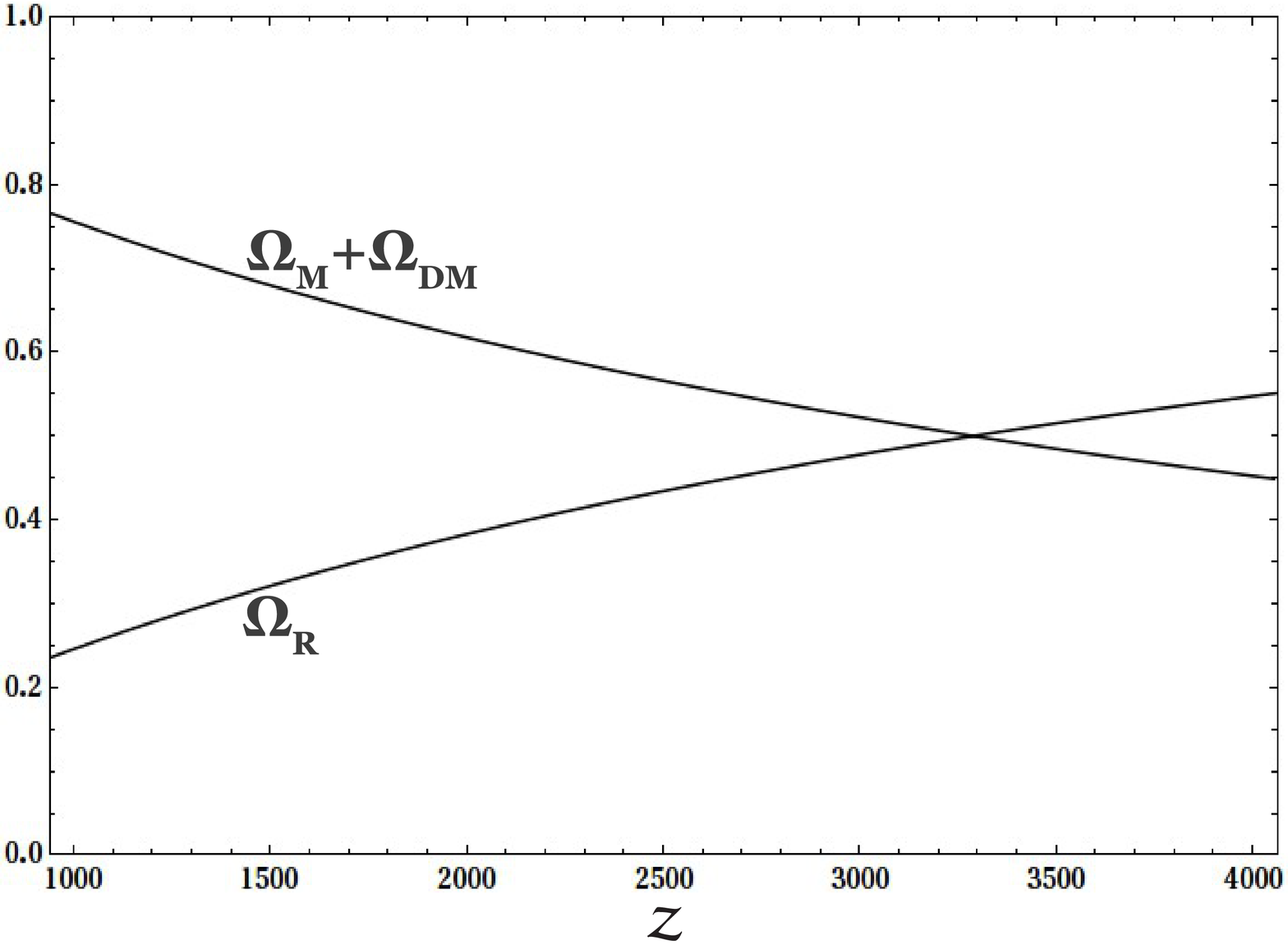}
\caption{Density parameters for $ 1000\leq z \leq 4000 $.}
\end{minipage}
\end{figure}

In Fig. 1 the density parameters are plotted as functions of the red-shift for values in the range $ 0\leq z \leq 1 $. From this figure we observe that the Universe is dominated by the dark energy, here described by the constant component of the fermionic field, which decreases as we go back in time. At a red-shift $ z\approx 0.45 $ the dark matter contribution begins to dominate the Universe. At this point the radiation does not have a representative contribution to the density parameter. The plot for large values of the red-shift is given in Fig. 2. The equality of the contribution of the matter and radiation fields occurs when $ z\approx 3250 $. This result is in good agreement with the parameterized observational data, since from reference \cite{WMAP9} we have $ z_{eq}=3265^{+106}_{-105} $. It is noteworthy that here we have taken into account the degrees of freedom of the relativistic neutrinos.
\begin{figure}[htbp]
    \centering
    \includegraphics[width=0.5\textwidth]{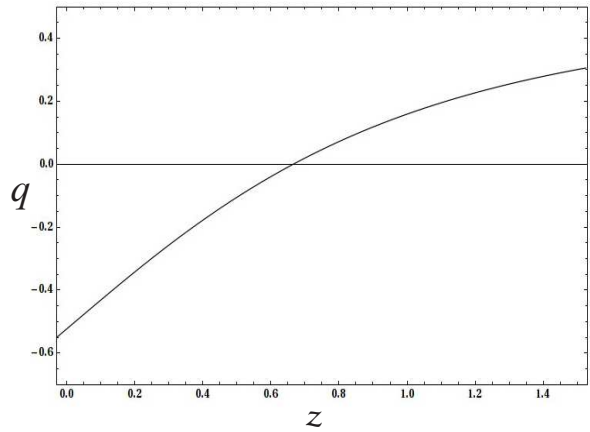}
    \caption{Deceleration parameter as a function of the red-shift.}
    \label{fig:last2}
\end{figure}

 The deceleration parameter as a function of the red-shift is represented in Fig. 3 and we observe from this graphic that the present value of the deceleration parameter is $ q(0)=-0.55 $, and the red-shift of the deceleration-acceleration transition is: $ z_{T}=0.67$. We have compared these values with the astronomical data in Table 1 (see \textit{e.g.} \cite{PhysRevD.86.083509}) and we infer that the obtained values of $ q_0 $ and $ z_T $  are in good agreement with the observational data.
\begin{table}
\begin{center}
 \begin{tabular}{|c|c|c|}
 \hline
    & Parametrization I & Parametrization II  \\
   $ q_0 $ & $ -0.61^{\tiny +0.06}_{\tiny -0.07} $ & $ -0.56^{\tiny +0.35}_{\tiny -0.22} $  \\
  $ z_T $ & $ 0.71^{\tiny +0.14}_{\tiny -0.17} $ &  $ 0.77^{\tiny +0.52}_{\tiny -0.57} $  \\ \hline
  &  Parametrization III & Fermionic Model \\
  $ q_0 $ &  $ -0.60\pm 0.06 $ & $ -0.55 $ \\
  $ z_T $ &  $ 0.72^{\tiny +0.27}_{\tiny -0.21} $ & $ 0.67 $ \\ \hline
    \end{tabular}
\end{center}
  \caption{Comparison of the deceleration parameter $ q_0 $ and the transition red-shift $z_T$ with the observational data (see \textit{e.g.} \cite{PhysRevD.86.083509})}
\end{table}

\subsubsection{The case $F'\neq0$:}

Let us search for a solution of the Friedmann equation (\ref{39}) when the self-interaction potential of the fermion field  and the non-minimal coupling function are given by
\ben
V=\Lambda+V_0\Psi=\Lambda+\frac{V_0\Psi_0}{a^3},\qquad F=\frac{1}{2}(1-\xi\Psi)=\frac{1}{2}\left(1-\frac{\xi\Psi_0}{a^3}\right),
\een
where $\xi$  is supposed to be a positive small coupling constant.

In this case the Friedmann equation can be rewritten as the following differential equation for the scale factor
\ben\lb{Fri}
3\left(a^3+2\xi\Psi_0\right)\dot a^2=\Lambda a^5 +\left(V_0\Psi_0+\rho_m^0\right){a^2}+{\rho_r^0}{a}.
\een

Three asymptotic cosmological solutions can be derived from the Friedmann equation (\ref{Fri}), namely,
\begin{enumerate}
  \item \emph{Primordial Universe in the presence of radiation:} in this case the terms $\Lambda$, $V_0\Psi_0$ and  $\rho_m^0$ can be neglected and (\ref{Fri}) becomes
  \ben\lb{s1}
  \left(a^2+\frac{2\xi\Psi_0}{a}\right)\dot a^2=\frac{\rho_r^0}{3}.\label{Approx}
  \een
  Two cases follow from the analysis of the above equation: (a) when the scale factor is small the approximated solution of (\ref{s1}) is
  \ben
  a(t)\simeq\frac{\rho_r^0}{24\xi\Psi_0}\left(t-t_0\right)^2,
  \een
  which describes an accelerated Universe characterized by a power-law inflationary period; (b) when the scale factor grows the non-minimal coupling dilutes and the solution of (\ref{s1}) reads
  \ben
  a(t)\simeq\sqrt{2}\left(\frac{\rho_r^0}{3}\right)^{1/4}\left(t-t_0\right)^{1/2},\label{RadEra}
  \een
  which refers to a radiation dominated period.
  \item \emph{Universe dominated by matter:} here the $\Lambda$ term can be neglected, the non-minimal coupling dilutes, the matter dominates the radiation  and (\ref{Fri}) reduces to
  \ben
  a\dot a^2=\frac{1}{3}\left(V_0\Psi_0+\rho_m^0\right),
  \een
  whose solution for the scale factor refers to the matter dominated period, namely,
  \ben
  a(t)\simeq\left({\frac{3}{2}}\right)^{2/3}\left(\frac{V_0\Psi_0+\rho_m^0}{3}\right)^{1/3}\left(t-t_0\right)^{2/3}.
  \een
  \item \emph{Present Universe:} since the radiation can be neglected in comparison to the other constituents and the non-minimal coupling is diluted, (\ref{Fri}) can be written as
  \ben
  a\dot a^2=\frac{1}{3}\left(\Lambda a^3 +V_0\Psi_0+\rho_m^0\right).
  \een
  The dilution of the non-minimal coupling implies that we are in the presence of the case analyzed in the last section where $F=1/2$, whose solution predicts the present decelerated-accelerated period, i.e., a matter dominated period goes into a de Sitter Universe in the future.

\end{enumerate}

\section{Hydrodynamical perturbations}

Let us now analyse the above solutions under small perturbations. In view of the difficulty of perturbing the spinor field through the Dirac equations, which are complicatedly coupled to the metric, we make an estimate of the perturbation dynamics through the hydrodynamic approach. To do this using the known method for the hydrodynamical perturbations, we will write the non-minimal model in a more adequate way, as follows.

By defining an effective energy-momentum tensor in the form $\widetilde{T}_{\mu\nu}={T_{\mu\nu}}/{2F}$, the non-minimal coupling model can be recast in the form of the Einstein's equations. Hence from action (\ref{1}) we have the generalized field equations which are recast in the form
\begin{equation}
R_{\mu\nu}-\frac{1}{2}g_{\mu\nu}R=-\frac{T_{\mu\nu}}{2F}=-\widetilde{T}_{\mu\nu}.
\end{equation}
So one has the Einstein's equations with the source of gravitational field being represented by the energy momentum-tensor $\widetilde{T}_{\mu\nu}$.

As a consequence, for a perfect fluid the canonical form of the energy-momentum tensor holds
\begin{equation}
\widetilde{T}_{\mu\nu}=\left(\widetilde{\rho}+\widetilde{p}\right)U_\mu U_\nu-\widetilde{p}g_{\mu\nu},
\end{equation}
where $U^\mu$ is the four-velocity and the effective energy density and pressure read
\begin{eqnarray}
\widetilde{\rho}=\frac{\rho}{2F}, \qquad \widetilde{p}=\frac{p}{2F}.\label{effFluid}
\end{eqnarray}
Thus the related Friedmann and acceleration equations are
\begin{eqnarray}
H^2=\frac{\widetilde{\rho}}{3}, \qquad \frac{\ddot a}{a}=-\frac{\widetilde{\rho}+3\widetilde{p}}{6},
\end{eqnarray}
and the conservation equation of the effective fluid has the usual form
\begin{equation}
\frac{d\widetilde{\rho}}{dt}+3H\left(\widetilde{\rho}+\widetilde{p}\right)=0.
\end{equation}

In this way, we can treat this hydrodynamical problem in the same way one does usually. The case for $F=1/2$ and $V=\Lambda+V_0\Psi$ produces the cosmological behavior of the models with baryonic and dark matter in the presence of a cosmological constant, whose perturbed solutions are known to be stable \cite{Mukhanov, Mukhanov1}. However, it is necessary to verify the stability against perturbations for the case with $F=(1-\xi\Psi)/2$ and $V=\Lambda+V_0\Psi$, when the non-minimal coupling presents some influence. This comprises the transition from the inflation to the radiation era whose approximated background solution was obtained from (\ref{Approx}), where the non-minimal coupling effect is present.

Since the background fluid is not anisotropic, the line element for small scalar perturbations in the longitudinal gauge \cite{Bardeen, Mukhanov, Mukhanov1} can be expressed as
\begin{equation}
ds^2=a^2\left[\left(1+2\Phi\right)d\eta^2-\left(1-2\Phi\right)\delta_{ij}dx^idx^j\right],
\end{equation}
where $\eta$ is the conformal time defined by $d\eta=dt/a$ and $\Phi$ is the gauge-invariant perturbation potential.

Once we are working in the framework of Einstein's equations with effective quantities, all the corresponding known results for the gauge-invariant scalar perturbations hold. But note that all the hydrodynamical quantities will refer to the effective ones above defined. Thus the perturbation potential equation for adiabatic perturbations is the usual one \cite{Mukhanov, Mukhanov1}
\begin{equation}
\Phi''+3\left(1+c_S^2\right)\mathcal{H}\Phi'-c_S^2\nabla^2\Phi+\left[2\mathcal{H}'+\left(1+3c_S^2\right)\mathcal{H}^2\right]\Phi=0,\label{PertPot}
\end{equation}
with the prime denoting derivative with respect to conformal time and $c_S$ is the speed of sound related to the effective fluid, $c_S^2=\partial \widetilde{p}/\partial\widetilde{\rho}$. The Hubble parameter is written in terms of the conformal time, $\mathcal{H}=a'/a^2$.

When the Universe is dominated by a fluid with barotropic state equation, $\rho=wp$, where $w$ is a positive constant, one has $c_S^2=w$ and the scale factor evolves as $a\propto \eta^{2/(1+3w)}$. For such a case, considering plane wave perturbations, $\Phi(\textbf{x}, \eta)=\Phi_{\textbf{k}}(\eta)e^{i\textbf{k}.\textbf{x}}$, equation (\ref{PertPot}) gives the solution
\begin{equation}
\Phi_{\textbf{k}}=\eta^{-n}\left[AJ_n\left(\sqrt{w}k\eta\right)+BY_n\left(\sqrt{w}k\eta\right)\right], \qquad n=\frac{1}{2}\left(\frac{5+3w}{1+3w}\right),\label{SolPert}
\end{equation}
where $A$, $B$ are constants and $J_n$, $Y_n$ are the Bessel functions of the first and second kind, respectively.

The barotropic equation of radiation is also valid in the effective form, $\widetilde{p}_r=\widetilde{\rho}_r/3$, as can be seen through (\ref{effFluid}). For $w=1/3$ one has $a\propto\eta$, which is the same approximated solution (\ref{RadEra}) in the presence of a non-minimal coupling sufficiently diluted.  The corresponding perturbation potential comes from (\ref{SolPert}) with $w=1/3$ and can be written in the following form
\begin{equation}
\Phi_{\textbf{k}}=\frac{1}{\widetilde{\eta}^2}\left[\left(\frac{A}{\widetilde{\eta}}+B\right)\sin{\widetilde{\eta}}+\left(\frac{B}{\widetilde{\eta}}-A\right)
\cos{\widetilde{\eta}}\right],\label{SolOne}
\end{equation}
where $\widetilde{\eta}=k\eta/\sqrt{3}$. Under the above considerations, this solution is a good approximation for the perturbations related to the asymptotic background solution (\ref{RadEra}).

Hence the time evolution of the energy density perturbations \cite{Mukhanov, Mukhanov1} of the effective fluid reads
\begin{eqnarray}
\frac{\delta\widetilde{\rho}_r}{\widetilde{\rho}_r}=2A\left[\left(\frac{2-\widetilde{\eta}^2}{\widetilde{\eta}^2}\right)
\left(\frac{\sin{\widetilde{\eta}}}{\widetilde{\eta}}-\cos\widetilde{\eta}\right)-\frac{\sin{\widetilde{\eta}}}{\widetilde{\eta}}\right]\nonumber\\
+4B\left[\left(\frac{1-\widetilde{\eta}^2}{\widetilde{\eta}^2}\right)
\left(\frac{\cos{\widetilde{\eta}}}{\widetilde{\eta}}+\sin\widetilde{\eta}\right)+\frac{\sin{\widetilde{\eta}}}{2}\right],\label{densPert}
\end{eqnarray}
where ${\delta\widetilde{\rho}_r}$ is a gauge-invariant perturbation. From (\ref{densPert}) we can assure that the amplitude of the effective energy density perturbations decays with time and asymptotically freezes out for large times.

Now we need to know if the amplitude of the original energy density perturbations, i.e. of $\rho_r$, also decays with time. Using our initial definition, $\widetilde{\rho}=\rho/2F$, we can write the effective density perturbations in terms of the original quantities, from which one obtains
\begin{equation}
\frac{\delta\widetilde{\rho}_r}{\widetilde{\rho}_r}=\frac{\delta{\rho}_r}{{\rho}_r}-\frac{\delta F}{F}.
\end{equation}

As the non-minimal coupling in question is $F=(1-\xi\Psi)/2$, one has
\begin{equation}
\frac{\delta\widetilde{\rho}_r}{\widetilde{\rho}_r}=\frac{\delta{\rho}_r}{{\rho}_r}+\xi\frac{\delta\Psi}{1-\xi\Psi}
\approx\frac{\delta{\rho}_r}{{\rho}_r}+\xi{\delta\Psi}.
\end{equation}
Note that this approximation holds because $F$ is sufficiently diluted and $\xi$ is very small.
From this expression we conclude that ${\delta{\rho}_r}/{{\rho}_r}$ as well as $\xi\delta\Psi$ must decay, since the corresponding sum gives a quantity that decays, i.e. ${\delta\widetilde{\rho}_r}/{\widetilde{\rho}_r}$. So the amplitude of the energy density perturbations of radiation and the $\Psi$ perturbations decay with time and asymptotically stabilize for large times. Thus, under the considered approximations, the non-minimal fermionic model in question appears to be stable against small perturbations.

\section{Final remarks and conclusions}

In this work we have considered a classical fermionic field minimally and non-minimally coupled with the gravitational field. The fermionic field is represented by a spinor field which is understood as a set of complex-valued space-time functions which transforms according to the Lorentz group. Classical spinors were discussed by Armend\'ariz-Pic\'on and Greene in the reference \cite{Piccongreene}. About such a consideration we point out that: (i) the expectation value of a spinor field in a physical state is a complex number and not a Grassmannian number, (ii) the spinor field can be treated classically if its state is close to the vacuum and (iii) we have extrapolated the validity of the classical spinor to the inflation despite a classical field theory could fail at the beginning of this regime.

\textit{Universe described by a fermion field} -- In the literature \cite{PhysRevD.69.124010, PhysRevD.74.124030, Marlosprd, Marlosepl} several forms for the potential density and coupling of the fermion fields were proposed in order to describe cosmological models with accelerated and decelerated periods. The  present work shows that, if the Noether symmetry is satisfied, the potential density and coupling have very restrictive forms.
The results for this case are: (i) the minimally coupled fermion field recovers the standard model for a Universe composed of matter and one has only a decelerated regime and (ii) in the non-minimal coupling with gravity the fermion field behaves as an inflaton.

\textit{Universe described by fermion, matter and radiation fields} -- When we impose the Noether symmetry  the minimal and non-minimal coupling cases admit a generic self interacting potential of the pseudo scalar. Here the results are: (i) the minimal coupling model reproduces a decelerated-accelerated regime, comprising the radiation era in the beginning and passing through the matter domination until the era when the dark energy dominates, which is described by the constant term of the fermionic potential and (ii) the non-minimal coupling can describe a Universe that begins with an accelerated expansion which goes into the radiation dominated era when the non-minimal coupling dilutes and after that it enters into the decelerated-accelerated period (matter-dark energy era).

In the literature \cite{rudicanonical, PhysRevD.79.063518, rudibosonfermion} several models with two different fields representing the dark energy and the dark matter were proposed in order to describe cosmological models with decelerated and accelerated periods. It is interesting that we obtained the same description with just one field representing either the dark matter and the dark energy, i.e., the fermionic field. To sum up, the Noether potentials and couplings of the model with their respective cosmological scenarios are displayed in Table 2.

\begin{table}
\centering
\begin{tabular}{|c|c|c|c|c|c|c|}
\hline
Cases & $V$ & $F$ & Inflation & Radiation era & Matter era & Current Accel  \\
\hline
 I & $\alpha\Psi$ & $\omega\Psi$ & $a\propto e^{\Lambda t}$ & --- & --- & --- \\
 \hline
 I & $\alpha\Psi$ & $\frac{1}{2}$ & --- & --- & Yes & --- \\
\hline
 II & $\Lambda+V_0\Psi$ & $\frac{1}{2}$ & --- & Yes & Yes & Yes \\
\hline
 II & $\Lambda+V_0\Psi$ & $\frac{1}{2}(1-\xi\Psi)$ & $a\propto t^2$ & Yes & Yes & Yes \\
\hline
\end{tabular}
\caption{Potentials and couplings and their cosmological scenarios.}
\end{table}

\section*{Acnowledgments}

G.M.K. acknowledges the financial support by CNPq and G.G. by CAPES.

\section*{Appendix. Noether symmetry condition}

Let us take a Lagrangian that does not depend explicitly on time, $\mathcal{L}=\mathcal{L}(q_i, \dot q_i)$, with the coordinates $q_i=q_i(t)$. The Euler-Lagrange equations of this Lagrangian read
\begin{equation}
\frac{\partial \mathcal{L}}{\partial q_i}-\frac{d}{dt}\left(\frac{\partial \mathcal{L}}{\partial \dot q_i}\right)=0.
\end{equation}
Consider now the following vector field
\begin{equation}
\textbf{X}=\alpha_i\frac{\partial}{\partial q_i}+\frac{d\alpha_i}{dt}\frac{\partial}{\partial \dot q_i},
\end{equation}
where the coefficients $\alpha_i$ are functions of the coordinates $q_i$. This vector field describes point transformations of the coordinates, $q_i\longrightarrow q'_i$, which induce velocity transformations,
\begin{equation}
q_i'=\frac{\partial q'_i}{\partial q_j}q_j \quad \Longrightarrow \quad \dot q_i'=\frac{\partial q'_i}{\partial q_j}\dot q_j.
\end{equation}
One assumes that these transformations are invertible. The quantity $\textbf{X}$ is also called infinitesimal generator of symmetry.

If the Lagrangian $\mathcal{L}$ is invariant under the point transformations represented by $\textbf{X}$, the following identity must hold
 \begin{equation}
L_{\textbf{X}}\mathcal{L}=\alpha_i\frac{\partial \mathcal{L}}{\partial q_i}+\frac{d\alpha_i}{dt}\frac{\partial\mathcal{L}}{\partial \dot q_i}=0,
\end{equation}
where $ L_{\textbf{X}}\mathcal{L}$ stands for the Lie derivative of $\mathcal{L}$ with respect to the vector field $\textbf{X}$.
Let us see the meaning of this invariance. If we contract the Euler-Lagrange equations of $\mathcal{L}$ with the coefficients $\alpha_i$, one obtains
\begin{eqnarray}
\alpha_i \frac{\partial \mathcal{L}}{\partial q_i}-\alpha_i\frac{d}{dt}\left(\frac{\partial \mathcal{L}}{\partial\dot q_i}\right)=0\\
\alpha_i\frac{\partial\mathcal{L}}{\partial q_i}-\left[\frac{d}{dt}\left(\alpha_i\frac{\partial\mathcal{L}}{\partial\dot q_i}\right)-\frac{d\alpha_i}{dt}\frac{\partial\mathcal{L}}{\partial\dot q_i}\right]=0\\
\frac{d}{dt}\left(\alpha_i\frac{\partial\mathcal{L}}{\partial\dot q_i}\right)=\alpha_i\frac{\partial\mathcal{L}}{\partial q_i}+\frac{d\alpha_i}{dt}\frac{\partial\mathcal{L}}{\partial\dot q_i}\equiv L_{\textbf{X}}\mathcal{L}.
\end{eqnarray}
Thus, since $L_{\textbf{X}}\mathcal{L}=0$ holds, it follows the Noether's theorem:\\

If $L_{\textbf{X}}\mathcal{L}=0$, there is a constant of motion given by $M_0=\alpha_i \frac{\partial \mathcal{L}}{\partial \dot q_i}$.\\

In other words, if the Lagrangian $\mathcal{L}$ is invariant under point transformations represented by $\textbf{X}$, there is a Noether symmetry associated with  $\mathcal{L}$.


\end{document}